\documentclass[aps,prl,twocolumn,showpacs,superscriptaddress,reprint]{revtex4-1}  

\usepackage{graphicx}
\usepackage{bm}
\usepackage{amssymb}
\usepackage{amsmath}
\usepackage{float}
\usepackage{pstricks} 
\usepackage{color}
\usepackage{bbold}
\usepackage{xr} 
\usepackage{tikz}
\usepackage{comment}
\usetikzlibrary{matrix}

\begin{document}

\title{Gaussian Optical Ising Machines}

\author{William~R.~Clements}
\email{william.clements@physics.ox.ac.uk}
\affiliation{Clarendon Laboratory, Department of Physics, University of Oxford, Oxford OX1 3PU, UK}

\author{Jelmer J. Renema}
\affiliation{Clarendon Laboratory, Department of Physics, University of Oxford, Oxford OX1 3PU, UK}

\author{Y. Henry Wen}
\affiliation{Clarendon Laboratory, Department of Physics, University of Oxford, Oxford OX1 3PU, UK}

\author{Helen M. Chrzanowski}
\affiliation{Clarendon Laboratory, Department of Physics, University of Oxford, Oxford OX1 3PU, UK}

\author{W. Steven Kolthammer}
\affiliation{Clarendon Laboratory, Department of Physics, University of Oxford, Oxford OX1 3PU, UK}

\author{Ian A. Walmsley}
\affiliation{Clarendon Laboratory, Department of Physics, University of Oxford, Oxford OX1 3PU, UK}

\date{\today}

\begin{abstract}{It has recently been shown that optical parametric oscillator (OPO) Ising machines, consisting of coupled optical pulses circulating in a cavity with parametric gain, can be used to probabilistically find low-energy states of Ising spin systems. In this work, we study optical Ising machines that operate under simplified Gaussian dynamics. We show that these dynamics are sufficient for reaching probabilities of success comparable to previous work. Based on this result, we propose modified optical Ising machines with simpler designs that do not use parametric gain yet achieve similar performance, thus suggesting a route to building much larger systems.}
\end{abstract}

\maketitle

\section*{Introduction}Combinatorial optimization problems, such as the travelling salesman problem, appear in many disciplines \cite{cook1998combinatorial}. However, finding an optimal solution to combinatorial problems is a hard task for conventional computers. Special-purpose hardware that can solve such problems more efficiently than conventional computers is therefore an active area of research. Examples include analog electrical circuits \cite{tank1986simple,yoshimura2015uncertain}, molecular computing \cite{adleman1994molecular,lipton1995dna}, and more recently adiabatic quantum computing \cite{kadowaki1998quantum,farhi2001quantum}.

One combinatorial optimization problem of particular note is the Ising problem \cite{barahona1982computational}. The Ising problem, in the absence of an external magnetic field, consists of finding the configuration of a network of coupled spins that minimizes the following Hamiltonian:

\begin{align}
H = -\sum_{i,j} J_{ij} \sigma_i \sigma_j
\label{Hamiltonian}
\end{align}

\noindent
where $\sigma_i$ and $\sigma_j$ are the values of the spins of sites $i$ and $j$ that can be either -1 or 1, and $J_{ij}$ is the $(i,j)$ entry of a matrix $\textbf{J}$ describing the spin-spin couplings. The Ising problem maps onto several physical and combinatorial problems, such as the maximum cut problem \cite{karp1972reducibility}. It is known to be NP-hard, but special-purpose hardware may find solutions to the problem faster than conventional computers. 

It has recently been suggested that a train of coupled optical pulses in a cavity undergoing parametric amplification could be used as such special-purpose hardware \cite{wang2013coherent,haribara2016coherent}. In their final state, these pulses oscillate as optical parametric oscillators (OPOs) with either a $0$ or a $\pi$ phase with respect to the pump light, and these two phases can be used to encode up or down spin directions. Coupling between pulses can be arranged in such a way that the system preferentially oscillates in a configuration that minimizes the Hamiltonian in equation \ref{Hamiltonian}. 

This coupling can be achieved in an optical delay line (ODL) architecture, where all the pulses coherently interfere with each other. ODL Ising machines have been experimentally demonstrated \cite{marandi2014network,inagaki2016large}, although with low connectivity between the encoded spins. Increasing connectivity and scale is an active area of research \cite{hp2017}. More recently, a new measurement and injection feedback (MIF) scheme, that implements the required coupling using partial homodyne measurements and electronic feedback, was demonstrated with up to two thousand coupled pulses \cite{inagaki2016coherent,mcmahon2016fully}. These devices were shown to outperform several other combinatorial optimization algorithms.

In light of their experimental demonstrations of OPO Ising machines, Inagaki \textit{et al.} \cite{inagaki2016coherent} and McMahon \textit{et al.} \cite{mcmahon2016fully}, raised interesting questions concerning the nature and role of quantum features and thresholding behaviour in the operation of their devices. Given the complex dynamics of the OPO and pump fields below and above threshold, elucidating the relevant computational mechanisms presents challenges. 

In this work we study MIF and ODL Ising machines that operate with simplified dynamics described by the Gaussian state formalism \cite{adesso2014continuous}. We show that these simplified dynamics are sufficient to attain high success probabilities similar to what has been experimentally demonstrated. Based on these results, we propose simplifications to experimental realizations of these devices, which we envisage will be of use to scaling them up to many more coupled pulses.



\section*{Theoretical Framework}

\begin{figure}[h!] 
\includegraphics[width=8.5cm]{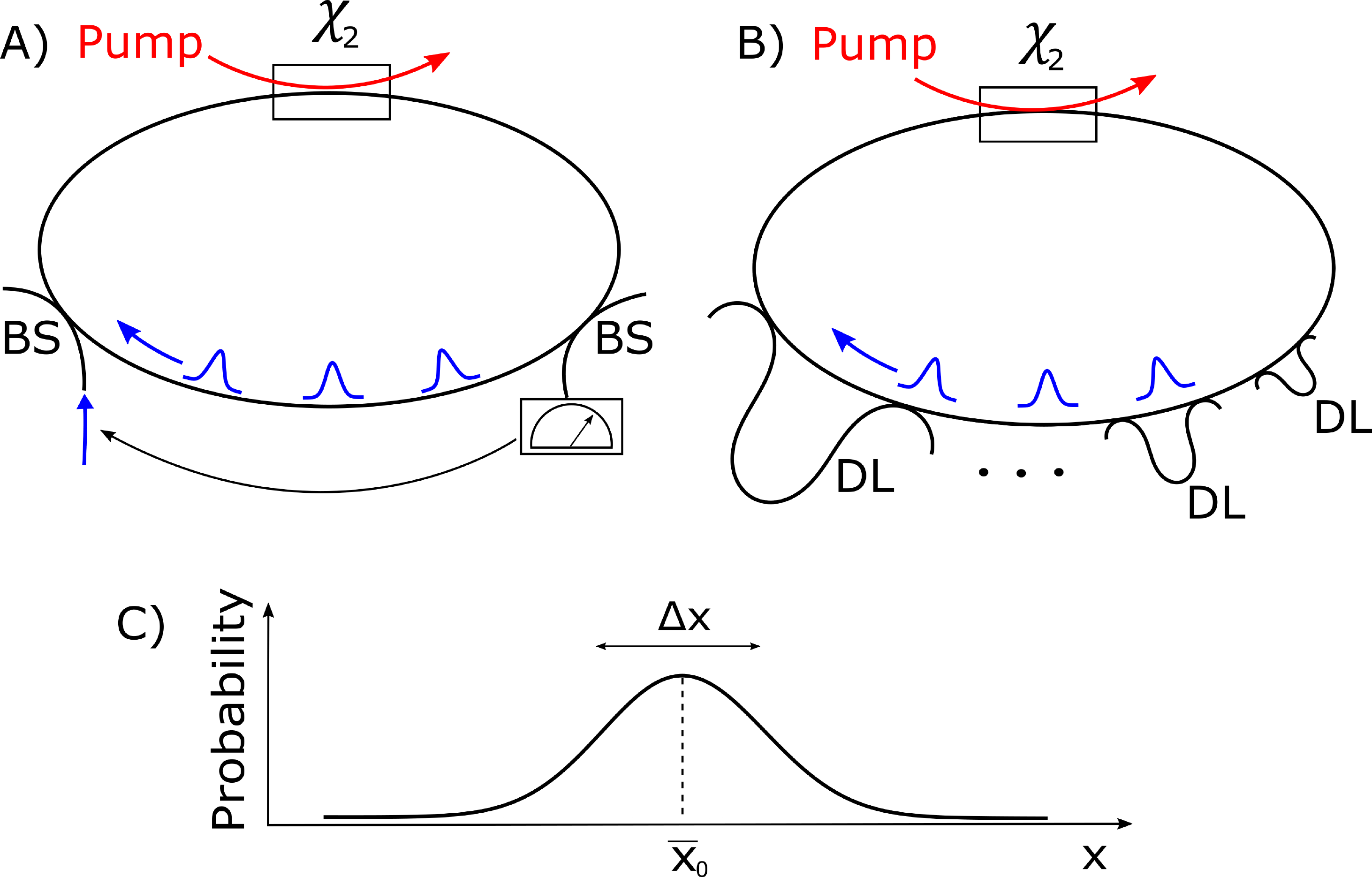}
\caption{Overview of optical Ising machines. A) In measurement and injection feedback Ising machines, in every loop optical pulses undergo parametric amplification in a nonlinear ($\chi_2$) material. A beam splitter (BS) picks off a fraction of each pulse for measurement by homodyne detection. The pulses are then displaced at a second BS by an amplitude dependent on the measurement result. B) In the optical delay line scheme, a part of each pulse is picked off into a delay line (DL) and interfered with a consecutive pulse. With a sufficient number of delay lines, every pulse can be made to couple to every other pulse. C) In Gaussian optical Ising machines, the optical pulses are described by a multidimensional Gaussian quasi-probability distribution in phase space, which is completely determined by its mean (the displacement) and by its covariance matrix.}
\label{fig1}
\end{figure}

Figure 1 provides an overview of the two different types of optical Ising machines considered in this work. In both cases, a train of pulses, initially in a vacuum state, circulates in a fiber loop with parametric gain. These pulses are coupled with each other via either MIF (Fig. 1a) or ODL (Fig. 1b). In the MIF scheme, this coupling is produced by splitting off a fraction of each pulse, upon which homodyne measurement is performed. Each pulse is then displaced in phase space via optical feedback, with an amplitude and phase determined by the collective measurement results of all the other pulses. In the ODL scheme, a fraction of each pulse is repeatedly picked off, delayed, and then interfered with a consecutive pulse. In both cases, the gain or the feedback is increased with each round trip and the system settles into a final configuration of steady state OPO pulses with well-defined phases. The sign of the phases of the pulses can then be measured and mapped onto spin orientations in an Ising model, in which the spin-spin coupling is determined by the optical coupling. The system preferentially settles into a configuration that corresponds to a low energy in the Ising model.

OPO Ising machines are usually described using the stochastic master equation formalism \cite{drummond1981non,maruo2016truncated}, which provides a picture of the interaction between the pump and the signal both below and above threshold. In this work, we consider optical Ising machines that are governed by simplified dynamics described by the Gaussian state formalism \cite{adesso2014continuous}. The Gaussian state formalism applies to optical states that have a Gaussian quasi-probability distribution in phase space, such as vacuum, coherent states, and squeezed states. An $N$-mode Gaussian state is fully described by its $2N \times 2N$ covariance matrix $\textbf{M}$ and $2N$-length displacement vector $\textbf{d}$ that characterize the Gaussian quasi-probability distribution. A Gaussian process maps a Gaussian state to another Gaussian state, and is described by simple matrix operations on the covariance matrix and displacement vector. 

We note that in OPO Ising machines, many physical processes are Gaussian processes. These include loss, beam splitting, displacements, and homodyne detection. Parametric amplification can also be approximated as a Gaussian squeezing operation on the signal mode as long as pump depletion is negligible, for example in an OPO operated below threshold \cite{milburn1981production}. A non-Gaussian model is necessary to account for the full interaction between the pump and the signal; this interaction is not considered in our work.

\section*{Gaussian Measurement-feedback Ising machines}

\begin{figure}[h!]
\includegraphics[width=8.3cm]{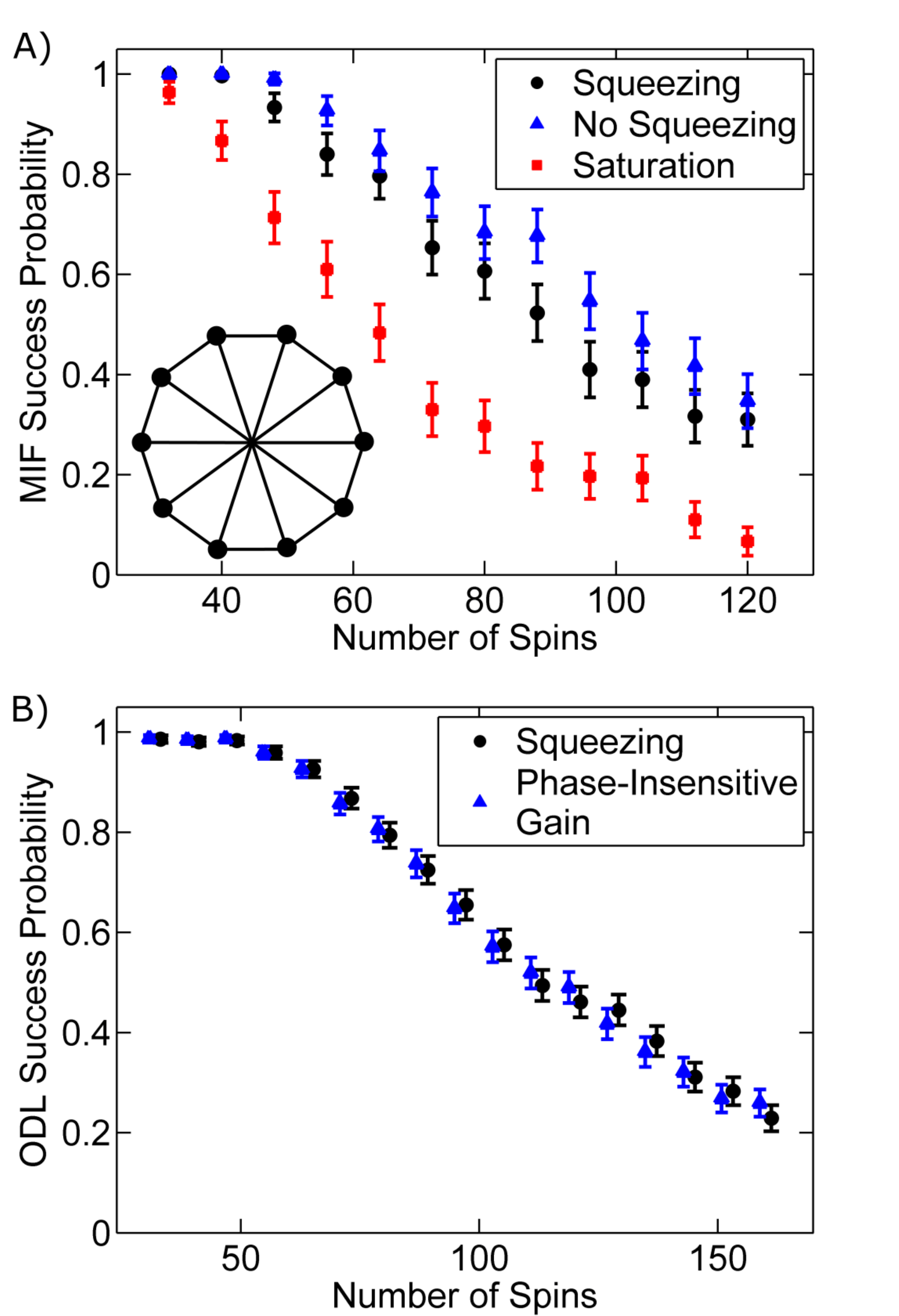}
\caption{Estimated probabilities of finding the correct ground state energies for M{\"o}bius ladder graphs of different sizes. A) Simulation of MIF Ising machines including squeezing (circles), without squeezing (triangles), and with saturated feedback (squares). The inset depicts a M{\"o}bius ladder graph. B) Simulations of ODL Ising machines with squeezing (circles) and a phase-insensitive gain medium (triangles). These data correspond to the same graph sizes, and are shown shifted for greater visibility. All error bars indicate a a 95\% confidence interval.}
\label{fig2}
\end{figure}

We simulate a Gaussian analogue of the MIF Ising machine demonstrated by McMahon \textit{et al.} \cite{mcmahon2016fully}. To provide a comparison of our results to their work, we calculate the success probabilities for finding the ground state energies of specific Ising systems, where the spin-spin couplings are given by M{\"o}bius ladder graphs (see figure \ref{fig2}a). The $(i,j)$ entry of the $\textbf{J}$ matrix corresponding to these M{\"o}bius ladders is $-1$ if nodes $i$ and $j$ are connected in the graph representing the M{\"o}bius ladder, and $0$ otherwise. McMahon \textit{et al.} found that their experimental system could find the correct ground state energy with a probability going from nearly 100\% for small systems to about 20\% for a 100-spin system.

Our simulation proceeds as follows. The pulses start in a vacuum state. In every loop, each pulse undergoes squeezing with a squeezing parameter of 0.2, followed by 30\% loss. We then pick off 10\% of each pulse on a beam splitter, on which we perform a simulated homodyne measurement. This is done by randomly drawing an array of numbers $\textbf{c}$ from the marginal Gaussian probability distribution determined by the covariance matrix and displacement vector of the measured light (see Methods). As described by McMahon \textit{et al.} we then displace the modes as follows:

\begin{align}
\textbf{d} &\rightarrow \textbf{d} + n \eta \textbf{J} \textbf{c}
\label{displace}
\end{align}

\noindent
where $n$ is the loop number and $\eta=0.001$ is the feedback strength. With these parameters, the pump power is initially roughly 25\% below the oscillation threshold of the system with $\eta=0$, which is similar to what was used in the experiment by McMahon \textit{et al.} We note that when $\eta>0$, the feedback term in equation \ref{displace} brings the system above threshold. 



After 300 loops through the system, we take the sign of the elements of $\textbf{d}$ to be the orientation of the corresponding spins, with which we calculate the energy of the spin configuration using equation \ref{Hamiltonian}. We repeat this process with graphs of up to 120 spins. The probabilities of finding the correct result for these different sizes, estimated from 300 trials for each size, are shown in figure 2a. We find a probability greater than $50\%$ of finding the correct result for sizes up to about 90 spins.

\begin{figure}[h!]
\includegraphics[width=8.5cm]{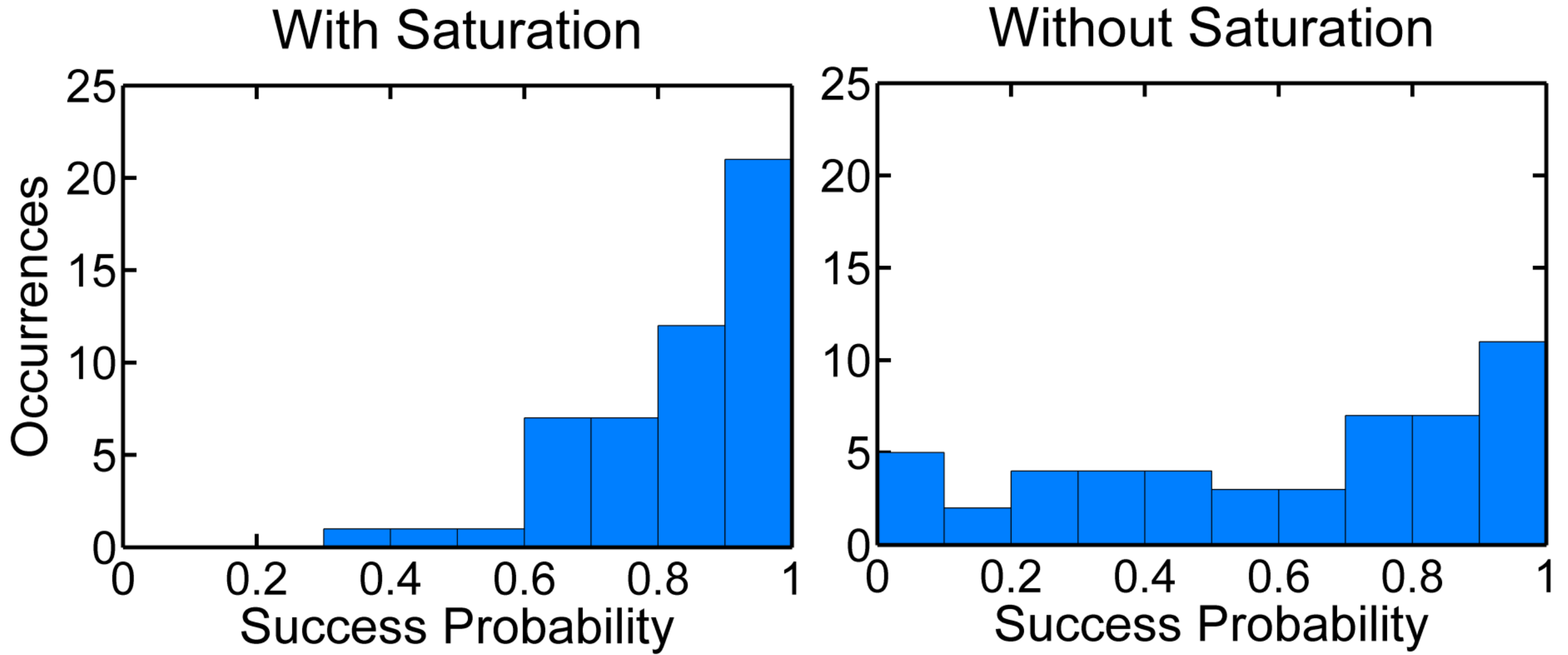}
\caption{Histograms of the success probabilities for 50 randomly selected cubic graphs with 16 spins, for a MIF Ising machine with (left) and without (right) saturated feedback. The probabilities are estimated from 100 simulation trials per graph.}
\label{fig3}
\end{figure}

The simulation above does not include gain saturation that would be present in a realistic machine, for example due to pump depletion, which could affect the success probability. To investigate the influence of saturation, we repeat our MIF simulations with the feedback displacement capped at 1000. In the case of a M{\"o}bius ladder graph the performance is adversely affected (Fig. 2A). However, we find that the change in performance is different for different graphs. Following McMahon \textit{et al.} \cite{mcmahon2016fully}, we calculate the success probability for 50 randomly chosen 16-spin cubic graphs, for which each spin is connected to exactly three others, using 100 trials per graph. Figure 3 compares our results for a saturated and an unsaturated MIF Ising machine. With saturation, low success probabilities are not observed. However, not every graph benefits. The success probability is only increased in 28 out of 50 cases. We observe similar outcomes for different values of the maximum feedback displacement, ranging from 100 to $10^6$.

\section*{Gaussian Optical Delay Line Ising machines}

We also study a Gaussian analogue to ODL Ising machines. In our simulation, the pulses circulate around the loop 100 times. In every round trip, each pulse undergoes squeezing with a squeezing parameter of 0.2. Then, for every pair of pulses $(i,j)$, if $J_{ij}$ is $-1$, 10\% of pulse $i$ is picked off on a beam splitter, given a $\pi$ phase shift, and interfered with pulse $j$ on a 90:10 beam splitter. After 100 round trips, we obtain a full description of the final optical state in phase space within the Gaussian state formalism. To simulate multiple experimental trials, we perform 1000 homodyne measurements on this state, from which we calculate an energy based on equation \ref{Hamiltonian}. We use these trials to estimate the probability of success (Fig. 2b). We find a success probability greater than $50\%$ for sizes up to about 110 spins, which is similar to what we found for the MIF scheme.

\section*{Interpretation}




Our simulation of Gaussian MIF Ising machines can be understood as implementing a classical discrete random walk in optical phase space, in which the position of the walker is specified by the displacement vector. Each step taken by the walker is directed by the random outcome of the homodyne measurement via the feedback. As the feedback strength increases over time, the walker tends to move towards a specific direction away from the origin. This direction corresponds to a spin configuration which minimizes the Ising energy. This process is reminiscent of other Monte Carlo optimization schemes such as simulated annealing \cite{kirkpatrick1983optimization}, in which a classical random walk probes parameter space.


In our simulation of Gaussian ODL Ising machines, we deterministically generate a probability distribution which is a multivariate Gaussian function. A specific spin configuration is found by drawing a sample from this distribution. This procedure is reminiscent of boson sampling with Gaussian states and Gaussian measurements \cite{aaronson2011computational}.

We note that both types of devices considered in this work can be efficiently emulated by classical computation. This can be seen as a consequence of the fact that Gaussian optical Ising machines involve only Gaussian measurements on Gaussian states \cite{bartlett2002efficient}.

\section*{Simplified Ising Machines}

The OPO Ising machines demonstrated by Inagaki \textit{et al} and McMahon \textit{et al} can potentially scale up to tens of thousands of coupled pulses \cite{haribara2016coherent}. Here we propose simplified Ising machines that may provide a route to building large scale devices.




We first consider a Gaussian MIF Ising machine in which we simply remove the squeezing but maintain the measurement and feedback. The pulses in the system are therefore coherent states. In the absence of the gain that previously partially compensated for the loss, we double the feedback strength to $\eta=0.002$. The random walk process implemented by these simulations follows the same pattern as in the simulation with parametric gain, but random numbers are now drawn from a narrower distribution around the average displacement due to the absence of squeezing. Even though the phase of a coherent state is not intrinsically bistable as for OPOs, the use of measurement and injection that displaces along only one quadrature axis effectively makes the phase bistable. Our simulation results are shown in figure 2a. We find that removing the squeezing has only minor influence on the success probability. We also find, as in the case with squeezing, that the success probabilities for different graphs can be modified by saturated feedback.


Another alternative to an OPO-based MIF Ising machine involves the use of phase insensitive gain media instead of parametric gain, as has been proposed \cite{utsunomiya2011mapping}. Such a scheme would allow for tuning the uncertainty on the quadratures of the pulses, thus adding an additional controllable degree of freedom to the random walk implemented by MIF Ising machines. 


We also simulate the Gaussian ODL Ising machine as before, except the squeezing operation is replaced with phase insensitive gain. We note that in this case, the phases of the pulses are no longer bistable, but they remain correlated due to their mutual couplings. Our results are shown in figure 2B. The performance with phase insensitive gain is essentially identical to that with squeezing.


\section*{Conclusion}

We introduce Gaussian optical Ising machines, and find that Gaussian dynamics are sufficient for finding low-energy solutions of the Ising problem, with performance comparable to demonstrated OPO based machines. Based on this finding, we propose simplified Gaussian approaches that suggest a route to developing future large scale optical Ising machines.


 
 
\vspace{0.5cm} 
 
We thank Jan Sperling, Alireza Marandi, Peter McMahon, Edwin Ng and Tatsuhiro Onodera for helpful discussions. We acknowledge support from NWO Rubicon, the European Research Council, the UK Engineering and Physical Sciences Research Council (project EP/K034480/1 and the Networked Quantum Information Technology Hub), from The John Templeton Foundation, and from the European Commission (H2020-FETPROACT-2014 grant QUCHIP).

\section*{Methods}

The following provides more detail as to how we implemented our simulations using the Gaussian state formalism. In our simulations, we used the convention that the covariance matrix of vacuum is $\frac{1}{2} \mathbb{1}$, but the following formulae do not depend on the choice of convention.

\subsection{Simulation of Squeezing and Loss}

Single mode squeezing on mode $i$ of a $N$-mode Gaussian state changes its covariance matrix $\textbf{M}$ and displacement vector $\textbf{d}$ as follows:

\begin{align}
\textbf{M} &\rightarrow  \textbf{S} \textbf{M} \textbf{S}^T \\
\textbf{d} &\rightarrow \textbf{S} \textbf{d}
\end{align}

\noindent
where $\textbf{S}$ is unity except for entries $2i-1$ and $2i$ along the diagonal which are equal to $e^r$, where $r$ is the (real) squeezing parameter.

Optical loss on mode $i$ is modelled by a beam splitter with transmission $t$ acting between mode $i$ and an ancilla vacuum mode. This changes $\textbf{M}$ and $\textbf{d}$ as follows:

\begin{align}
\textbf{M} &\rightarrow  \textbf{T} \textbf{M} \textbf{T}^T +  \textbf{R} \textbf{V} \textbf{R}^T \\
\textbf{d} &\rightarrow \textbf{T} \textbf{d}
\end{align}

\noindent
where $\textbf{T}$ is unity except for entries $2i-1$ and $2i$ along the diagonal which are equal to $\sqrt{t}$, $\textbf{R}$ is unity except for entries $2i-1$ and $2i$ along the diagonal which are equal to $\sqrt{1-t}$, and $\textbf{V}$ is the covariance matrix for an $N$-mode vacuum. 

\subsection{Simulation of Homodyne Measurements}

A homodyne measurement on the $N$-th optical mode of an $N$-mode Gaussian state described by covariance matrix $\textbf{M}$ and displacement vector $\textbf{d}$ is simulated as follows. $\textbf{M}$ can be written in block-diagonal form:

\begin{align}
\begin{pmatrix}
    \textbf{A} & \textbf{B}\\
    \textbf{B}^T & \textbf{C}
  \end{pmatrix}
\end{align}

\noindent
where $\textbf{C}$ is the reduced $2 \times 2$ covariance matrix for the homodyned mode, $\textbf{A}$ is the reduced covariance matrix for the remaining $N-1$ modes, and $\textbf{B}$ keeps track of the correlations between these modes. The homodyne measurement projects mode $N$ onto an infinitely squeezed state with a displacement vector $\textbf{c}$, which is drawn from the projection of the marginal Gaussian probability distribution described by $\textbf{C}$ and $\textbf{d}$ onto the chosen measurement axis. The remaining $N-1$ modes are then described by a new covariance matrix $\textbf{A'}$ and displacement vector $\textbf{d'}$:

\begin{align}
\textbf{A'} &= \textbf{A} - \textbf{B}(\textbf{P}\textbf{C}\textbf{P})^+\textbf{B}^T \\
\textbf{d'} &= \textbf{d}_A + \textbf{B}(\textbf{P}\textbf{C}\textbf{P})^+(\textbf{c}-\textbf{d}_C)
\end{align}

\noindent
where $\textbf{P}$ is a projector onto the measured quadrature of mode $N$, $^+$ denotes the Moore-Penrose pseudo-inverse, and $\textbf{d}_A$ and $\textbf{d}_C$ are the original reduced displacement vectors for the first $N-1$ modes and for the $N$-th mode, respectively.

\subsection{Simulation of Phase Insensitive Gain}

Phase-insensitive gain on a given mode $i$ of an $N$-mode state can be modelled in the Gaussian state formalism by using an ancilla vacuum mode, implementing a two-mode squeezing operation with squeezing parameter $r$ on these two modes, and tracing out the ancilla mode. In the $(x_1,p_1,...,x_N,p_N)$ basis in optical phase space, the covariance matrix $\textbf{M}$ and the displacement vector $\textbf{d}$ for the amplified mode are thus transformed as follows:

\begin{align}
\textbf{M} &\rightarrow  \textbf{C} \textbf{M} \textbf{C}^T + \textbf{S}\textbf{V}\textbf{S}^T\\
\textbf{d} &\rightarrow \textbf{C} \textbf{d}
\end{align}

\noindent
where $\textbf{C}$ is unity except for entries $2i-1$ and $2i$ along the diagonal which are equal to $\cosh(r)$, $\textbf{S}$ is unity except for entries $2i-1$ and $2i$ along the diagonal which are equal to $\sinh(r)$, and $\textbf{V}$ is the covariance matrix for an $N$-mode vacuum. In our simulation of an ODL Ising machine with phase-insensitive gain, we used a squeezing parameter of $r=0.6$, which yields an amount of gain $\cosh(r) \approx 1.19$ that is similar to the gain $e^{r_1} \approx 1.22$ provided by single mode squeezing with $r_1=0.2$.

\bibliographystyle{ieeetr}
\bibliography{referencesIsingSimulations}

\end{document}